# The Size of Software Projects Developed by Mexican Companies

J. Aguilar[1], M. Sánchez[2], C. Fernández[2], E. Rocha[2], D. Martínez[2], and J. Figueroa[2]
[1]Departament of Engineering, Universidad Popular Autónoma del Estado de Puebla, Puebla, Pue., México
[2]Institute of Computer Engineering, Universidad Tecnológica de la Mixteca , Huajuapan, Oax., México

**Abstract -** *Currently, most software projects around the world are small rather than large. Despite this, there are more methodologies, tools, frameworks, processes, and so on, for developing and managing large software projects than for small ones. Small software projects are important because they generate considerable resources. For example: apps (small mobile applications) generate around $25 billion dollars of revenue. This paper shows our findings regarding the size of the projects built by Mexican software development companies. We surveyed 107 Mexican companies and found that 92% of their developed projects are micro and small, and 8% are medium or large. In addition, according to our research, 84.1% of companies in Mexico are micro or small businesses.*

## 1 Introduction

There are many examples of tools, methodologies or, processes oriented to medium or large projects such as the following: CMMI[13], COCOMO [14], EPM [15], among others. These tools, methodologies and processes are highly accepted around the world. We need the same global acceptance for small software projects, tools, methodologies and processes, because micro and small software projects generate an important economic impact. For example, apps are mainly small projects, and the app stores run by Apple Inc, Google Inc, and so on, offer more than 700,000 apps each and, generate around $25 billion dollars of global revenue [7].

First of all, we need to identify the size of software projects. There are several factors, among them: i) Project development cost, ii) Number of people required to develop the software, and iii) Amount of software to be produced [6]. For example, a small software project has these factors: i) Cost between USD 7,500.00 and 192,000.00; ii) Team of less than 4 persons; and, iii) Size between 9 KLOCs and 38 KLOCs. This is our personal viewpoint about small software projects, although each author has his or her own viewpoint about software size classification.

In this paper we will show the findings of our research regarding: a) Size of software projects being developed in Mexico, and b) Company size in Mexico. We want to answer the following questions: What is the size of software projects developed by micro and small software companies in Mexico? How is software size to be classified? These questions are important to answer because there are currently many software process models oriented to large and medium projects, but micro and small projects have perhaps been forgotten.

## 2 Background

This section shows the main elements of our research.

### 2.1 Software Size Classification

First of all, we are going to show different viewpoints regarding software size classification. Fred Brooks [1] said that there are different kinds of software: i) Programs complete in themselves, ready to be run by the author on the system on which it was developed, ii) A programming product that can be run, tested, repaired and extended by anybody, iii) A Programming System, which is a collection of interacting programs, and iv) Programming Systems Product, which costs nine times as much as the other three kinds. Another author, Watts Humphrey [2], said that there are five stages of Software Product Size: Stage 0, very small program elements, written by programmers alone. Stage 1, small programs, or modules, designed, implemented, and tested by programmers alone; these programs typically range in size from only a few dozen to several hundred LOCs (Lines of Code). Stage 2, larger programs, or components, that typically involve teams of developers who develop and integrate multiple stage-1 modules into larger stage-2 component programs. Stage 3, very large projects that involve multiple teams controlled and directed by a central project management. Finally, Stage-4, massive multi-systems that involve many autonomous or loosely federated projects. The next author, Solvita Berzisa [3], said that project size is described with four attributes where three attributes –team size, budget and duration- are scalar values and such values should be grouped into intervals. The team size values have been divided into two intervals: less than seven and seven and more. Analyzed budget attribute values have been divided into five intervals: less than 10 000 USD, 10,000 to 50,000 USD, 50,000 to 100,000 USD, 100,000 to 500,000 USD, and more that 500,000 USD. Project duration values have been divided into four intervals: less than six months, six months to a year, from a year to two years and more than two years. These authors did their own software size classification.

Besides these viewpoints, The International Software Benchmarking Standards Group (ISBSG) is the global and independent source of data and analysis for the IT industry. There are 5,052 projects included in the ISBSG[1] in June 2009, of which 76.1% are micro and small projects, from 0 to 399 function points, which corresponds to approximately 31,920 LOCs. We applied Backfiring[4] to convert function points to lines of code. Smaller projects are more common in this repository. The projects have been submitted from 24 countries. Projects of different size ranges have different key characteristics [5][8][9]. For example, we can use table 2.1 to find the ideal duration and staffing ranges for a project, based on its estimated effort.

**Table 2.1** Time and effort in software projects according to their size.

| Project Size | Effort. Work hours of effort. | Months Duration | Ideal staffing |
|---|---|---|---|
| Small | 8-360 | 0-3 | 1 person |
| Medium | 361-3600 | 3-6 | 3-7 persons |
| Large | 3601-24 000 | 6-12 | 7-24 full-time persons |

A software project can be considered to be small if one or more of the following criteria apply:

a) Less than two man-years of development effort is needed.
b) A single development team of five people or less is required.
c) The amount of source code is less than 10 KLOCs.

The size of a particular project can vary greatly depending on the language chosen. For instance a 200,000 line Perl project generally has the functionality of a 1,200,000 line C project. This is from the estimate in Code Complete that states that Perl usually requires 1/6 the number of lines to do the same task as C. However, different size ranges entail different issues that involve major organizational differences. Specifically:

1. Small - reasonable size for one person to produce in a reasonable time.

2. Medium - reasonable size for a small team (max 8; beyond that size, communication issues arise that drop productivity and require reorganization to handle) to develop and entirely understand.
3. Large - A large team (significantly > 20 - note that the average team of 20 is about as productive as a team of 5-8 due to organizational issues).
4. Huge - The project is large enough that specialized tools are probably needed to help with project navigation so that team members can figure out what they need to understand to work on it.

As can be seen, there are a lot of viewpoints regarding software size classification. In this paper we try to take into account all these opinions and offer a proposal.

## 2.2 Economic impact of small software projects.

Basically, an app is a small software project. For instance, the average size of an Apple app is around 10 KLOCs [12]. For this paper, then, we consider an app as a small software project. Thus, we can see that there are millions of mobile devices sold around the world. For example, in 2012, there were 446 million Android devices, 199 million Apple devices and, 17 million Windows devices sold. Furthermore, there are millions of apps available [11]. The apps market was about USD 25 billion which is expected to be USD 155 billion in 2017 [10]. Obviously, the economic impact of small software projects is very important.

## 3 Our research

This section describes in detail our research; we only show our results about what kind of software are mainly implemented by Mexican software development companies. Section 3.1 will explain the methodology used for our work. Section 3.2 describes the results of our research, explaining the software size classification, company size classification and the size of Mexican software projects.

### 3.1 Methodology description

We began this research by doing a systematic search of the literature related to our project. So far, we have not found similar studies at least for Mexican companies. It is important to mention that this study is part of a larger project in which we want to identify good techniques and practices for developing small projects. The problem is that there are not enough tools, processes, methodologies and so on for small software projects. This problem was identified by some of our researchers, because they had been working directly in software development companies.

For this project, we wanted to have a better knowledge of Mexican companies working in software development projects and to gather data about characteristics of their small software projects. In order to get this data, we contacted software companies from Mexico and offered a free SCRUM course to members from participating companies. We received answers from 107 Mexican companies. The results of our research are detailed in section 3.2.

---

[1] The global and independent source of data and analysis for the IT industry. www.isbsg.org

## 3.2 Results

We analyzed data gathered from our survey, where 107 Mexican companies were asked about: a) Their size, and b) The size of the projects they developed. We found that 92% of their projects were micro and small projects, and 8% were medium or large projects. Additionally, we found 84.1% are micro and small companies and they barely use processes, methodologies or tools for developing and managing their small projects.

The survey was mainly focused on the size of projects being developed in Mexico and company size. We wanted to answer the question, what is the size of software projects being developed by micro and small software companies in Mexico?.

### 3.2.1 Kind of Projects Developed

With our research, we could answer this question: Q1. Since your company was founded, what percentage of projects (micro, small, medium, large) has your company developed? The companies were able to answer this question because we gave them our Project Size Classification Table (Table 3.2.1) in which they saw our classification of the sizes of different projects.

**Table 3.2.1** Project Size classification

| Project Size | Features |
|---|---|
| Micro | • Size (Lines of source code): from 1,600 to 9,600<br>• Development time (months): from 1 to 3<br>• Development time (hours): from 160 to 960<br>• Team size (members): from 1 to 2<br>• Cost (USD): 1,200 to 4,800* |
| Small | • Size (Lines of source code): from 9,601 to 38,400<br>• Development time (months): from 3 to 6<br>• Development time (hours): from 961 to 3,840<br>• Team size (members): from 2 to 4<br>• Cost (USD): 7,501 to 192,000* |
| Medium | • Size (Lines of source code): from 38,401 to 960,000<br>• Development time (months): from 7 to 60<br>• Development time (hours): from 3,841 to 96,000<br>• Team size (members): from 5 to 10<br>• Cost (USD): from 192,050 to 4,800,00.00* |
| Large/Big | • Size (Lines of source code): from 960,000 to ∞<br>• Development time (months): from 61 to ∞<br>• Development time (hours): from 96,001 to ∞<br>• Team size (members): from 11 to ∞<br>• Cost (USD): from 4,800,00.00 to ∞* |
| Note | *The cost is different in each country. We took USD 50.00 per man-hour. |

Our data analysis showed us than Mexico is a country where mainly micro & small software projects are developed, consisting of 92% of software projects (Fig. 3.2.2.1). Medium & large software projects represent a much smaller percentage-- only 8%. We can see these results in figure 3.2.2.1.

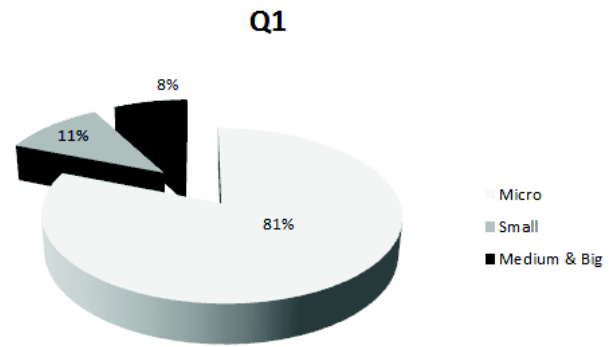

**Figure 3.2.2.1** Software projects developed in Mexico

We could also answer another question: Q2: What was the cost of your small projects?

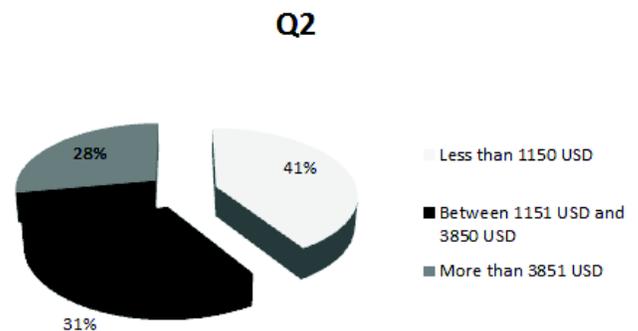

**Figure 3.2.2.2** Small software projects cost

The main activity of Mexican companies is to develop small projects, and the price of these small projects was mainly very low. 72% cost less than $3,850 USD ($ 50,000.00 Mexican pesos) (Fig. 3.2.2.2). Thus, we can infer: Mexican software development companies have developed small projects and its costs have been very low. Therefore, we need to give them mainly information to develop and manage micro & small projects.

Additionally, we posed this question: Q3: How long was needed to develop a small project?

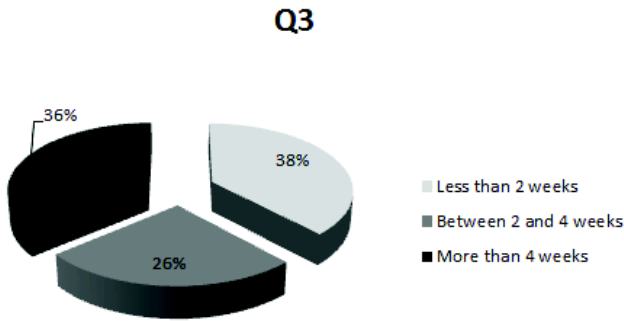

**Figure 3.2.2.3** Time to develop small software projects

We can see that Mexican software development companies need only a few weeks to develop a small project. 64% of projects required only one month or less to finish (Fig. 3.2.2.3). The question Q2 showed us the low cost to develop small projects. It may be that the reason for this low cost is the relatively little time necessary to develop this kind of project.

### 3.2.2 Company Size Classification.

We could answer another question: Q4. How many people work in your company?. The companies had four options: a) 1-5, b) 6-10, c) 11-50, d) 51 or more. When we analyzed the answers, we found that 84.1% were micro & small companies and 15.9% were medium and big (large) (table 3.2.2).

**Table 3.2.2 Q2,** Mexican Companies: Size classification.

| Size | Employees | Number of Mexican companies. | Percentage |
|---|---|---|---|
| Micro | 1-10 | 62 | 57.9% |
| Small | 11-50 | 28 | 26.2% |
| Medium & Big (Large) | 51 or more | 17 | 15.9% |
| Total | 107 | 107 | 100% |

This table shows us that in Mexico there are more micro and small companies than medium and big or large (Fig. 3.2.2.4). Thus, we need to focus on developing tools, processes, methodologies, model life cycles, and so on, for the micro and small sector.

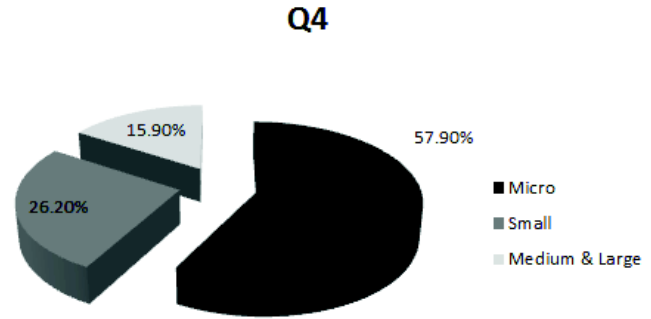

**Figure 3.2.2.4** Mexican companies' size

## 4 Future Work

With our findings, we have established the situation regarding the size of software developed and companies' size in Mexico. Based on these results, we will need to research the following: What kind of tools, methodologies and processes are the companies currently using? Which are most useful for Mexican companies? What must a Mexican company do to successfully adopt a tool, methodology or process to support small projects' administration and development? Perhaps, we will need to make some proposals for new tools, new process or new methodologies focusing mainly on our Mexican context.

## 5 Conclusion

We have found useful information to develop a strategic plan to help Mexican companies develop quality small software projects. For example, we have found: a) Most of the companies surveyed have developed a considerably high number of small projects. We can thus infer that the main activity of these companies is to develop small software projects. b) The cost to develop small software projects is very low; this could be a problem when the companies need economic resources to buy software or hardware for their operations. c) The time required to develop small projects is only one month or less, although 36% of small projects needed more than one month to be finished. d) Mexico is a country where 84% of software development countries are micro or small. We have not shown what percentage of small projects fail, but in this research we can see: e) Failure percentage in small projects is directly related to software-company size; the reasons for increasing failure in small projects are not clear. Finally we can say: In Mexico, mainly, software development companies need: tools, processes and methodologies to develop small software projects. The small software projects have an important economic impact. It is very important to be aware of the size of companies in our country or region, and to know the size of the projects developed by them, in order to make plans to support them.